\documentclass{PoS}
\usepackage[noadjust]{cite}
\usepackage{amsmath}
\usepackage{amssymb}

\title{Atmospheric neutrino spectrum reconstruction with JUNO}

\ShortTitle{Atm. $\nu$ spectrum reconstruction with JUNO}

\author{\speaker{Giulio Settanta}\thanks{Corresponding author}, Stefano M. Mari, Cristina Martellini, Paolo Montini, on behalf of the JUNO Collaboration\\
        Universit\`a degli Studi Roma 3 \& INFN, sezione di Roma 3\\
        E-mail: \email{giulio.settanta@uniroma3.it}}

\abstract{The atmospheric neutrino flux represents a continuous source that can be exploited to infer properties about Cosmic Rays and neutrino oscillation physics. The JUNO observatory, a 20\,kt liquid scintillator currently under construction in China, will be able to detect atmospheric neutrinos    , given the large fiducial volume and the excellent energy resolution. The light produced in neutrino interactions will be collected by a double-system of photosensors: about 18.000 20" PMTs and about 25.000 3" PMTs. The rock overburden above the experimental hall is around 700 m and the experiment is expected to complete construction in 2021.\\
In this study, the JUNO performances in reconstructing the atmospheric neutrino spectrum have been evaluated. The different time evolution of scintillation light on the PMTs allows to discriminate the flavor of the primary neutrinos. To reconstruct the time pattern of events, the signals from 3" PMTs only have been used, because of the small time resolution. A probabilistic unfolding method has been used, in order to infer the primary neutrino energy spectrum by looking at the detector output. The simulated spectrum has been reconstructed between 100\,MeV and 10\,GeV, showing a great potential of the detector in the atmospheric low energy region. The uncertainties on the final flux, including both statistic and the systematic contributions, range between 10\% and 25\%, with the best performances obtained at the GeV.}

\FullConference{
European Physical Society Conference on High Energy Physics - EPS-HEP2019 -\\
			10-17 July, 2019\\
			Ghent, Belgium}

\begin{document}

\section{Introduction}
\noindent The origin and properties of Cosmic Rays (CRs) are still matter of study by a number of experiments. A fraction of the energy of an air shower, resulting after a CR interaction in the atmosphere, is carried by neutrinos. The atmospheric $\nu$ flux is almost entirely composed of ${\nu_\mu}$ and ${\nu_e}$\footnote{Here, and in all the text, in $\nu_\mu$ and $\nu_e$ we include also the corresponding antineutrinos.} and spans many decades in energy, from the MeV up to the PeV scale. Neutrinos travelling across the Earth can be also used as a probe to study flavor oscillation effects in matter. After the experimental results in past years \cite{atmoFrejus,atmoAMANDA,atmoANTARES,atmoSK,atmoIC3,atmoIC4}, further contributions to the atmospheric $\nu$ spectrum will come from the next generation of neutrino detectors, operating in the next decade. JUNO is a large liquid scintillator (LS) detector with low energy threshold and low energy resolution ($\sim 3\%/\sqrt{E~[MeV]}$), currently under construction in China \cite{YBJuno}. Its core consists of a $\sim$36 m diameter acrylic sphere, which will be filled with 20\,kt of liquid scintillator. The light produced in $\nu$ interactions will be collected by two independent systems of photosensors: sabout 18.000 20" PMTs and 25.000 3" PMTs, respectively. Given the large fiducial volume and the excellent energy resolution, JUNO will be able to detect atmospheric neutrinos. The expected event rate is of the order of several $\nu$ interactions/day. The central sphere is surrounded by a cylindrical water pool, which acts as an active veto against cosmic muons. The cherenkov light produced inside the pool will be observed by about 2.000 20" PMTs. At the top of the system another detector, made of plastic scintillating strips, will be assembled: the top tracker, whose role is to identify and precisely track cosmic muons.

\section{The Monte Carlo simulation}
\begin{figure}[h!]
\centering
\includegraphics[width=0.66\textwidth]{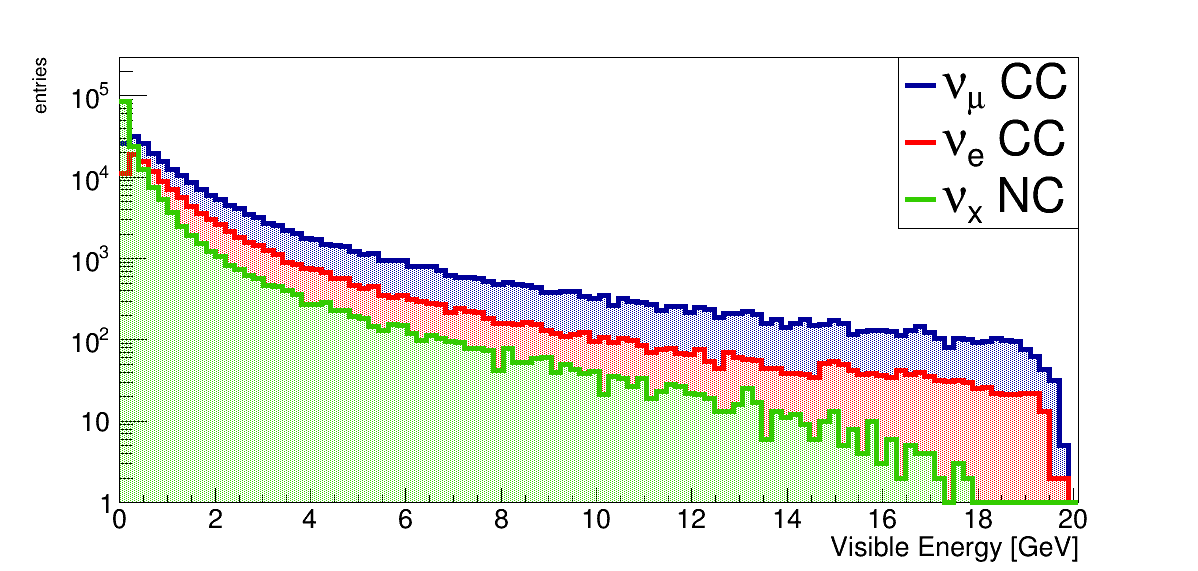}
\includegraphics[width=0.33\textwidth]{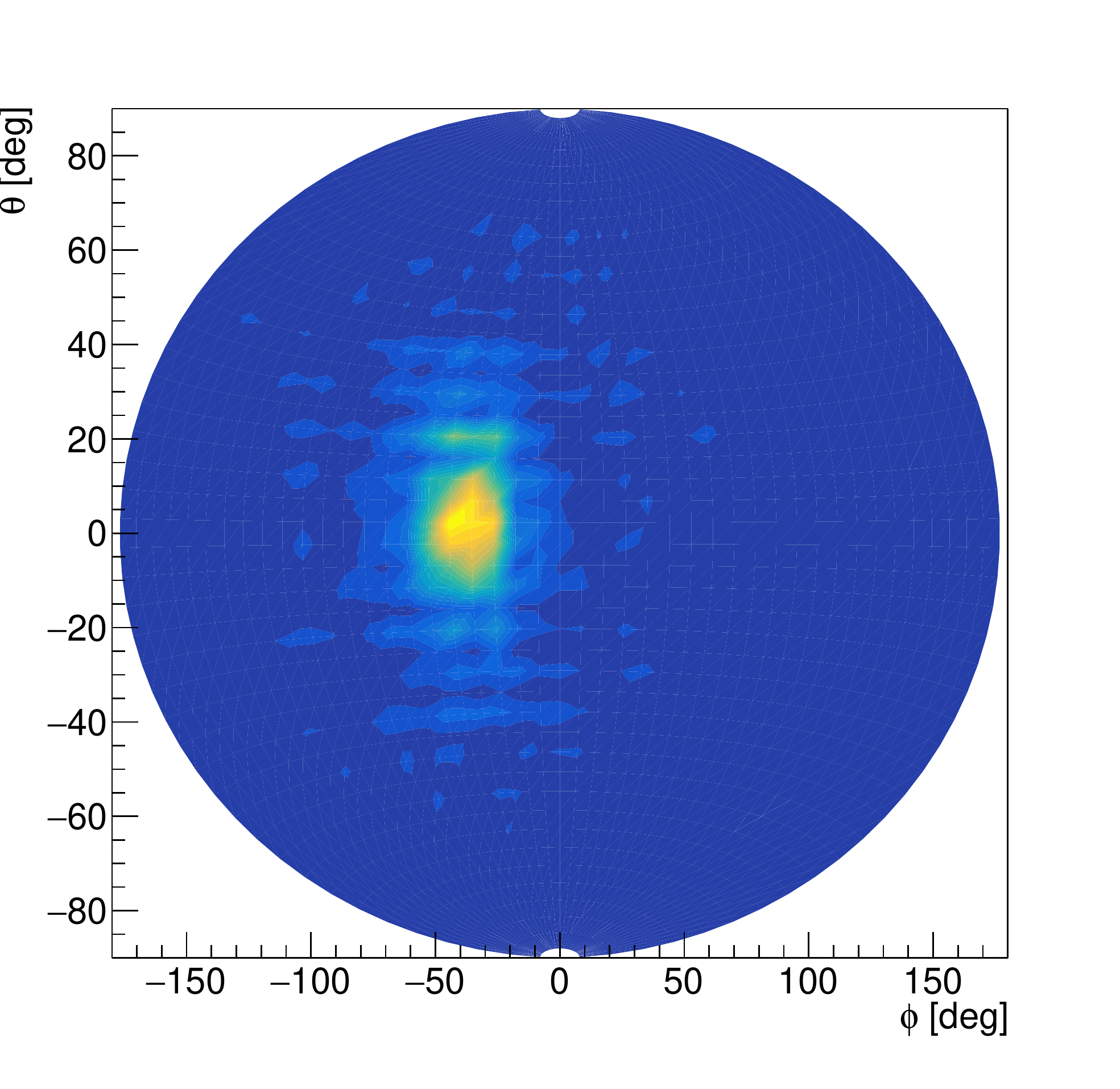}
\caption{Left: visible energy of the MC sample, for $\nu_\mu$ CC (blue), $\nu_e$ CC (red) and NC events (green); right: map of the hits on the 3" PMT system after one neutrino event ($E_\nu \simeq$ 900\,MeV, $\nu_e$ CC interaction).}
\label{fig:MC-1}       
\end{figure}

\noindent Since the JUNO detector is still under construction, the study relies on Monte Carlo (MC) simulations only. The MC production has been processed in two main steps:
\begin{itemize}
    \item Neutrino interactions generation within the detector volume. An initial flux of atmospheric $\nu$ at the JUNO location has been assumed, according to the predictions from \cite{hkkm15}. The model provides also the zenith- and azimuth-angle flux dependence and is calculated at the source. In order to get a realistic flux at the detector, flavor oscillation effects have been applied to the original flux, including both the vacuum and the matter contribution.\\
    The interaction of neutrinos with the target materials has been managed by the \emph{GENIE Neutrino Monte Carlo Generator} \cite{GENIE}. The interactions have been generated with a neutrino energy distribution up to 20 GeV and consist of $5 \cdot 10^5~ {\nu_\mu} + {\nu_e}$ events (fig. \ref{fig:MC-1} - left). The output of the simulation is the full list of secondaries and their associated properties (ID, momentum, direction $\dots$), including the type of interaction neutrinos undergo, either a charged-current (CC) or a neutral-current (NC) one.
    \item Propagation of secondaries in the scintillator by means of a \emph{GEANT4-}based simulation \cite{Sniper}, which includes energy losses, photon production and propagation, photon-PMT cathode interaction and photo-electrons (PEs) generation. Because of the JUNO very large active volume and the thousands of PMTs, running a full simulation is both CPU and storage consuming, which puts a limitation in the number of simulated events.
\end{itemize}
After the end of the simulation chain, a map of the hits on the detector is available, distributed in time and position (fig. \ref{fig:MC-1} - right).

\section{Sample selection and flavor identification}
\begin{figure}[h!]
\centering
\includegraphics[width=\textwidth]{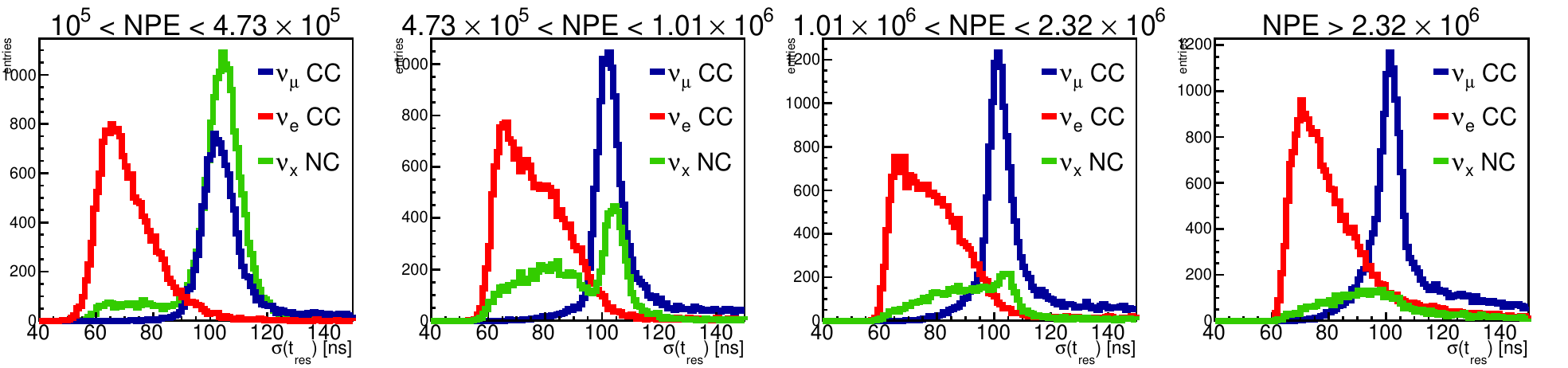}
\caption{Distribution of $\sigma(t_{res})$ for $\nu_\mu$ CC (blue), $\nu_e$ CC (red) and NC events (green). The NPE range is reported above the figures.}
\label{fig:tres}
\end{figure}

\noindent Simulated atmospheric neutrino events are selected in order to get two separate samples, one enriched in $\nu_e$ and the other one in $\nu_\mu$. Being a large LS, the JUNO central detector is not the ideal instrument for particle tracking. Indeed, its best performances are achieved in events which are entirely contained within the volume, where a full calorimetric measurement can be performed, with no need to separate the single tracks. Fiducial cuts applied to the neutrino sample are targeted to select events whose secondaries start and end within the CD, here called fully-contained (FC). On the contrary, events which have some secondary particles escaping from the CD are defined partially-contained (PC) and will have poorer energy resolution. Muons that may come from neutrino interactions outside the detector are not considered in the study. On the other hand, the above criteria form an intrinsic limitation for the spectrum reconstruction of high-energy $\nu_\mu$, namely $E_{\nu_\mu} \gtrsim$10\,GeV, because the resulting muon from a CC interaction is never fully contained inside the CD.\\
In order to select FC well-reconstructed events and enrich the sample in CC interactions, a series of fiducial cuts have been applied:
\begin{description}
    \item[$\mathbf{R_{VERTEX}<16\,m}.$] The cut on the vertex distance from the centre of the sphere $R_{VERTEX}$ has been used to remove events which release their energy near the edge of the acrylic sphere. These events tipically exhibit a loss of linearity between the deposited and the collected energy, because part of the energy is released in the acrylic and not in the LS and because the closest PMTs collect a great amount of light and can undergo saturation. In order to reproduce the uncertainty on the reconstructed vertex position, the true MC position has been smeared by means of a 3D gaussian function, centered at the true position and with the square root of the variance $\sigma$ = 1\,m.
    \item[$\mathbf{NPE_{LPMT}^{CD}>10^5}.$] To remove the core of NC events, which accumulates at low energy (fig. \ref{fig:MC-1} - left), a cut on the total charge collected by the 20" PMT system in the inner detector has been employed (NPE $\equiv$ Number of Photo-Electrons). $10^5$ PEs roughly correspond to a visible energy of 80\,MeV.
    \item[$\mathbf{NPE^{WP}<60}.$] The cut on the total charge collected in the water pool has been used to remove cosmic muon events and PC neutrino events.
    \end{description}
The overall efficiency of fiducial cuts on the neutrino sample is about 60\%. After the selection, the neutrino sample is composed at 97\% of FC events. The remaining PC events are composed at 96\% of $\nu_\mu$ CC interactions. To estimate the residual contamination coming from atmospheric muons, a full MC simulation of millions of events would be necessary, since the muon flux is several orders of magnitude higher than the atmospheric neutrino flux. However, since these kind of simulations are extremely CPU-consuming, a simplified 2D toy model has been implemented. The model simulates all the physical processes for light production and detection, including the stochastic fluctuations. As a validation, the model is able to reproduce the expected performances of the water pool veto (98\% efficiency) if a simple majority is requested to tag cosmic muon events. The estimated rejection power is $< 1.15 \cdot 10^{-6}$ @ 90\% confidence level, after the fiducial cuts.\\
A crucial part in the atmospheric $\nu$ study is the correct identification of the original $\nu$ flavor. Above $\sim$100 MeV neutrinos can be assumed to interact with nucleons only. This implies that hadronic particles are always produced in the final state. In the case of a CC interaction, the corresponding charged lepton is also produced (either $\nu_\mu~N \rightarrow \mu~X$ or $\nu_e~N \rightarrow e~X$). In the case of a NC interaction, only secondary hadrons and the scattered neutrino are in the final state. CC interactions are therefore the preferred channel for the neutrino flavor identification, since the lepton immediately identifies the original neutrino. Specifically, muons with energy > 1 GeV travel in general for a longer distance inside the detector with respect to electrons. Additionally, muons can decay inside the scintillator volume, giving a delayed energy release at low energy too. The above differences make $\nu_\mu$ CC events more elongated in time with respect of $\nu_e$ CC events, which indeed appear more point-like. Hadronic particles, which make the totality of a NC event, have in general a long-living energy release, because of interactions and decays. Furthermore, since hadrons are included in every neutrino-nucleon interaction, they make more similar $\nu_\mu$ and $\nu_e$ CC event topologies.\\
The different temporal behaviour of the classes of events has been exploited by building a time profile-based discrimination algorithm. The 3" PMT system only has been used in the algorithm, since it is more accurate in time measurement. To reproduce realistic Time-Transit Spread (TTS) effects, an artificial smearing has been applied to the true hit time over every PMT. A gaussian function with $\sigma = 4~ns$ has been used. A hit time-residual is then defined for each SPMT as:
\begin{equation}\label{eq:tres}
t_{res}^i = t_{hit}^i - \left( \frac{n \cdot R_V^i}{c} \right),
\end{equation}
where $t_{hit}^i$ is the hit time on the i-th SPMT, $c/n$ is the speed of light inside the scintillator and $R_V^i$ is the distance between the i-th SPMT and the interaction vertex. The smeared vertex position is again used. The time residual is therefore the arrival time of light on the PMTs, minus the time of flight of photons from the vertex to the PMT and reflects how smeared in time is the event.\\
Since $\nu_\mu$ and $\nu_e$ CC events result in different light production duration, the RMS of the $t_{res}$ distribution is used as a discrimination variable (here called $\sigma(t_{res})$). In fig. \ref{fig:tres} the $\sigma(t_{res})$ distribution is reported for the three populations: $\nu_\mu$ CC, $\nu_e$ CC and NC events. The variable is also reported separately in 4 different bins of NPE collected by CD LPMTs, selected in order to have equal statistics in each bin.

\noindent A good isolation of the $\nu_e$ CC component is evident, over the whole energy range. On the contrary, the $\nu_\mu$ CC and the NC components appear to be overlapped. The hadronic part of the secondaries has indeed an important contribution from charged pions, which decays with $\sim$100\% probability into $\mu + \nu$ and are therefore an irreducible background. Also protons and neutrons have timing features similar to muons, because they scatter for a long time before stopping. The NC component contribution, however, becomes less significant at high energy, because of the steeper spectral shape (fig. \ref{fig:MC-1} - left). Two separate cuts have been therefore applied for $\nu_e$ and $\nu_\mu$, to isolate CC events. The residual NC events, which have not been removed by the $t_{res}$ selection, are populated both of $\nu_e$ and $\nu_\mu$.\\
Concerning the $\nu_e$ selection, a value of $\sigma(t_{res}) <$ 75\,ns is required. The cut results in an efficiency about 41\% with respect to the sample after fiducial cuts and a residual contamination from $\nu_\mu$ less than 6\%. For the $\nu_\mu$ selection, a requirement of $\sigma(t_{res}) >$ 95\,ns is coupled to a further requirement of NPE > 500.000. The efficiency is 60\% with respect to the sample after fiducial cuts. The residual $\nu_e$ contamination is less than 20\%.

\section{Spectrum reconstruction}
\begin{figure}[h!]
\centering
\includegraphics[width=0.45\textwidth]{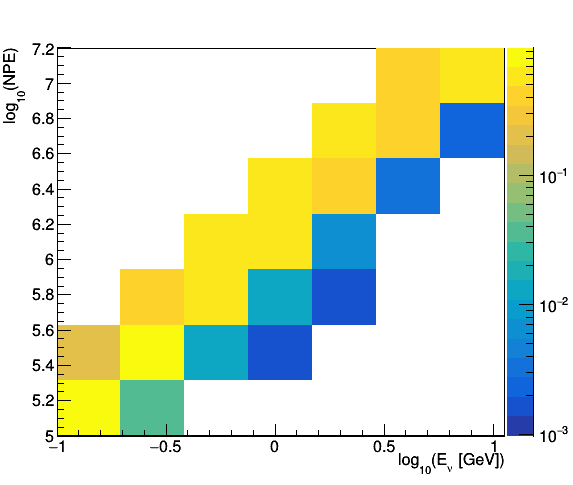}
\includegraphics[width=0.45\textwidth]{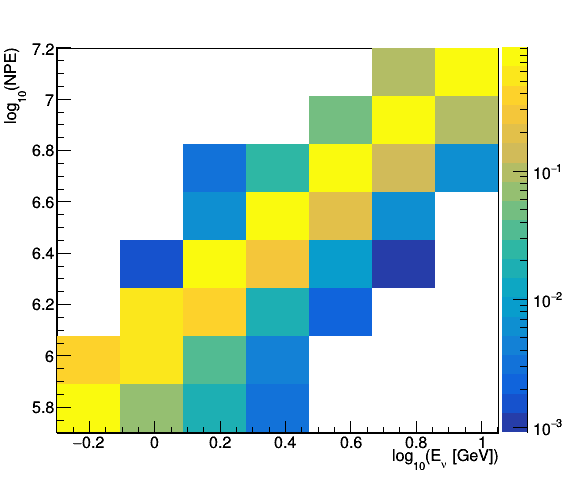}
\caption{Unfolding matrices for the $\nu_e$ (left) and the $\nu_\mu$ flux (right).}
\label{fig:UM}       
\end{figure}

\noindent In order to extract the primary neutrino energy spectrum, an unfolding procedure has been used. The observable spectrum $M_j$, which in this case is the NPE$^{CD}_{LPMT}$ distribution, is the result of:
\begin{equation}
M_{j}=\sum_{i} A_{j i} N_{i},
\end{equation}
where $N_i$ is the unfolded neutrino spectrum and $A_{ji}$ is the detector response matrix, which can also be expressed as the conditional probability that a neutrino with a certain energy produces a certain amount of NPE: $A_{j i}=P\left(N P E^{CD}_{L P M T_j} | E_{v_{i}}\right)$. The relationship can be inverted and the primary spectrum can be expressed as:
\begin{equation}
N_{i}=\sum_{j} \boldsymbol{U}_{i j} \boldsymbol{M}_{j},
\end{equation}
where $U_{ij}$ is the unfolding matrix. To build $U$, an iterative Bayesian method has been adopted \cite{GDA1,GDA2}. The detector response matrix has been estimated using the full MC sample and then normalized as $\sum_{j} A_{j i}=1-\epsilon$, where $\epsilon$ takes into account the inefficiency due to the reduced phase space considered by the matrix. As a prior flux distribution, the prediction from \cite{hkkm15} has been used. The two unfolding matrices, for the $\nu_e$ and the $\nu_\mu$ flux, are reported in fig. \ref{fig:UM}. As for the $A$ matrix, the unfolding matrix can be expressed as the conditional probability that an event which produced a certain amount of NPE originated from a neutrino with a certain energy: $U_{ij}=P\left(E_{\nu_i} | N P E^{CD}_{L P M T_j}\right)$: $U$ allows therefore to unfold the experimentally observed NPE distribution into the primary neutrino flux.

\begin{figure}[h!]
\centering
\includegraphics[width=0.49\textwidth]{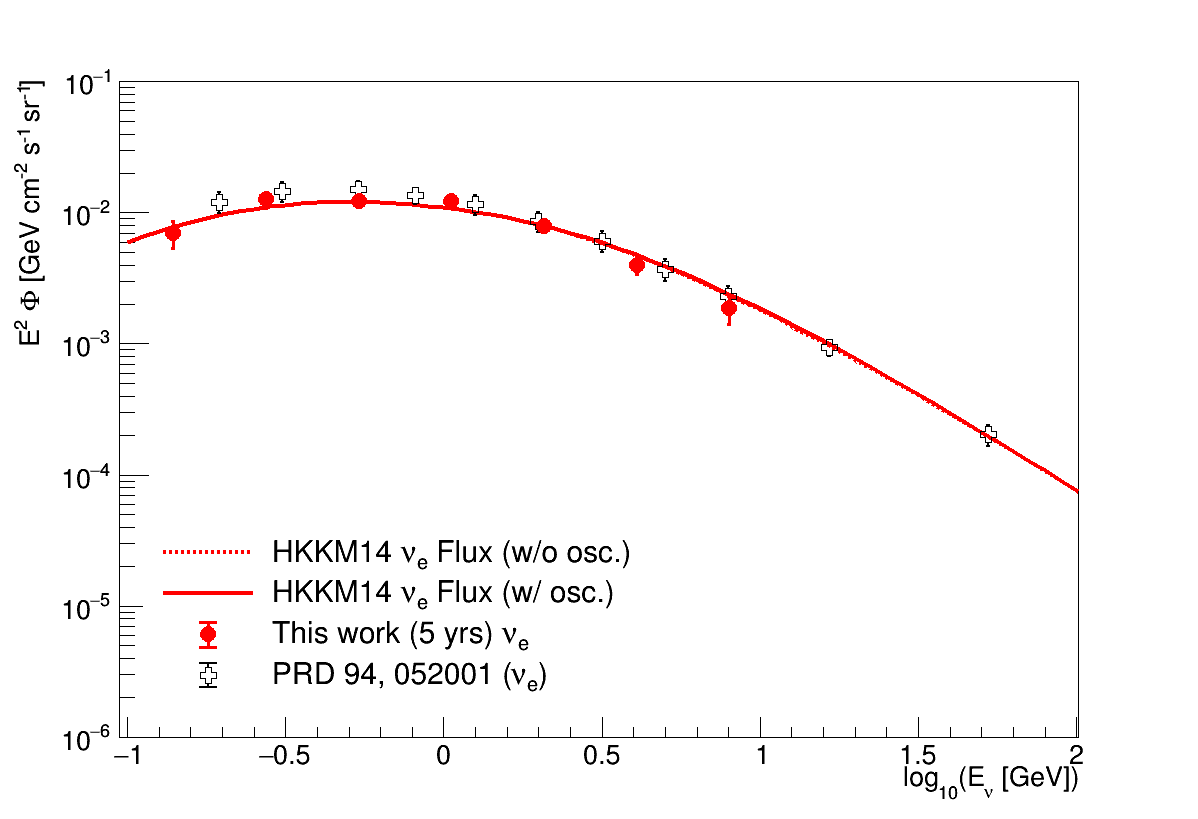}
\includegraphics[width=0.49\textwidth]{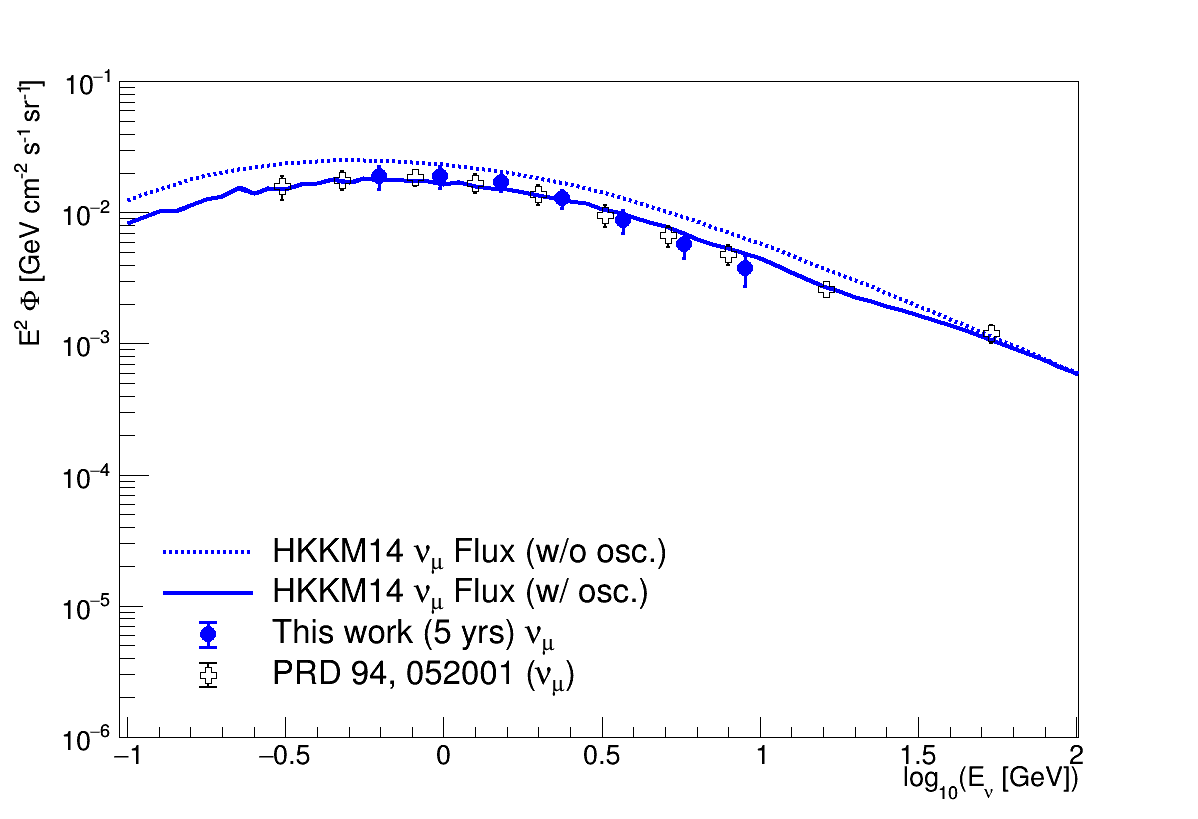}
\caption{Reconstructed atmospheric neutrino spectrum for $\nu_e$ (left) and $\nu_\mu$ (right). The prediction from \cite{hkkm15} is also shown, w/ (full line) and w/o (dashed line) the oscillation effects. The measurement from \cite{atmoSK} is reported (white crosses).}
\label{fig:spec}       
\end{figure}

\noindent An independent MC sample, corresponding to about 5 years of detector livetime, have then be generated, to simulate real data. The primary neutrino spectrum has been reconstructed separately for $\nu_e$ and $\nu_\mu$ and is reported in fig. \ref{fig:spec}, together with the prediction from \cite{hkkm15} and the measurement presented in \cite{atmoSK}.

\noindent The uncertainties reported in the spectra in fig. \ref{fig:spec} include both the statistic and the systematic contribution. In the second class, the effects from oscillations parameters and cross sections uncertainty, sample selection and probability matrices building have been included. The total uncertainty ranges between 10\% and 25\%, with the best performances obtained around the GeV.

\section{Conclusions}
\noindent The JUNO experiment, once completed, will be the largest liquid scintillator-based detector ever built. Among the many physics goals that JUNO is able to achieve, the atmospheric neutrino sector represents a target that can be addressed in the first years of data-taking. The large detector volume and the fine energy resolution allow the measurement of the atmospheric neutrino flux in the energy range [100\,MeV - 10\,GeV], with an uncertainty < 25\%. An event time-profile algorithm has been used for the $\nu_e$/$\nu_\mu$ separation. The flux measurement will provide information from an interesting energy region, where the oscillation effects are maximal and theoretical models have significative uncertainties.

\end{document}